\documentclass{forArxiv}
\usepackage{url}
\usepackage[numbers]{natbib}

\usepackage[english]{babel}
\usepackage[version=4]{mhchem}
\usepackage{physics}
\usepackage{textgreek}

\usepackage{silence}

\usepackage{xcolor}

\newcommand\ri{\right}
\renewcommand\le{\left}
\newcommand\la{\langle}
\newcommand\ra{\rangle}

\jname{Xxxx. Xxx. Xxx. Xxx.}
\jvol{AA}
\jyear{YYYY}
\doi{10.1146/((please add article doi))}

\begin{document}

\markboth{Joshi et al.}{Joshi et al.}

\title{Emergent Spatiotemporal Organization in Stochastic Intracellular Transport Dynamics}

\author{Kunaal Joshi,$^{1,\ast}$ Harrison York,$^{2,\ast}$ Charles S. Wright,$^{1,2}$ Rudro R. Biswas,$^1$ Senthil Arumugam$^{2,3,4,5,\dagger}$, and Srividya Iyer-Biswas,$^{1,6,\dagger}$
\affil{$^1$Department of Physics and Astronomy, Purdue University, West Lafayette, IN 47907, USA}
\affil{$^2$Monash Biomedicine Discovery Institute, Faculty of Medicine, Nursing and Health Sciences, Monash University, Clayton/Melbourne, VIC 3800, Australia}
\affil{$^3$ARC Centre of Excellence in Advanced Molecular Imaging, Monash University, Clayton/Melbourne, VIC 3800, Australia}
\affil{$^4$European Molecular Biological Laboratory Australia (EMBL Australia), Monash University, Clayton4/Melbourne, VIC 3800, Australia}
\affil{$^5$Single Molecule Science, University of New South Wales, Sydney, NSW 2052, Australia}
\affil{$^6$Santa Fe Institute, Santa Fe, NM 87501, USA}
\affil{$^\ast$These authors contributed equally to this work.} 
\affil{$^\dagger$To whom correspondence should be addressed: senthil.arumugam@monash.edu and iyerbiswas@purdue.edu.} 
}

\WarningFilter{xcolor}{Incompatible color definition}
\WarningFilter{latex}{Overful}
\WarningFilter{latex}{Underful}

\begin{abstract}
The interior of a living cell is an active, fluctuating, and crowded environment. Yet, it maintains a high level of coherent organization, which is readily apparent in the intracellular transport network. Membrane-bound compartments called endosomes play a key role in carrying cargo, in conjunction with myriad components including cargo adaptor proteins, membrane sculptors, motor proteins, and the cytoskeleton. These components coordinate to effectively navigate the crowded cell interior and transport cargo to specific intracellular locations, even though the underlying protein interactions and enzymatic reactions exhibit stochastic behavior. A major challenge is to measure, analyze, and understand how, despite the inherent stochasticity of the constituent processes, the collective outcomes show an emergent spatiotemporal order that is precise and robust. This review focuses on this intriguing dichotomy, providing insights into the known mechanisms of noise suppression and noise utilization in intracellular transport processes, and also identifies opportunities for future inquiry.
\end{abstract}

\maketitle


\section{INTRODUCTION}

Biological functions and chemical reactions within eukaryotic cells are spatially restricted and compartmentalized in both membrane-bound and membrane-less organelles. To reach specific destinations within the cell, molecules and organelles rely on cues for guidance. While these processes display consistency at a macroscopic level, they are intrinsically stochastic due to probabilistic elements at the subcellular level. Thermal fluctuations at the molecular scale influence diffusion, molecule binding, and reaction kinetics. Low numbers of components can lead to significant fluctuations relative to the mean. Even genetically identical cells can exhibit stochastic variations in protein copy numbers, which can be amplified in domains with limited binding capacity like lipid membranes. Interestingly---and importantly---this noise is not debilitating in complex biological processes that involve multiple components with diverse interactions, such as intracellular trafficking.

This review focuses on the interplay between constituent stochastic dynamics and deterministic outcomes in cellular organization. We describe (1) the components and organization of the vesicular transport network, (2) the physical and biochemical processes that govern cargo delivery within the cell, (3) examples of emergent trafficking processes that ensure robust transport outcomes, and (4) methods amenable to the study of fast, stochastic transport processes over sufficiently long periods. Finally, we discuss the future prospects of studying stochastic transport phenomena via whole-cell measurements and integrating imaging, mathematics, and biology to uncover underlying mechanisms.

\section{ORGANIZATION OF THE ENDOSOMAL TRANSPORT NETWORK}\label{sec:organization}

\begin{figure}[h]
  \includegraphics[width=5.75in,page=1,trim={0in 0.2in 0in 0.8in},clip]{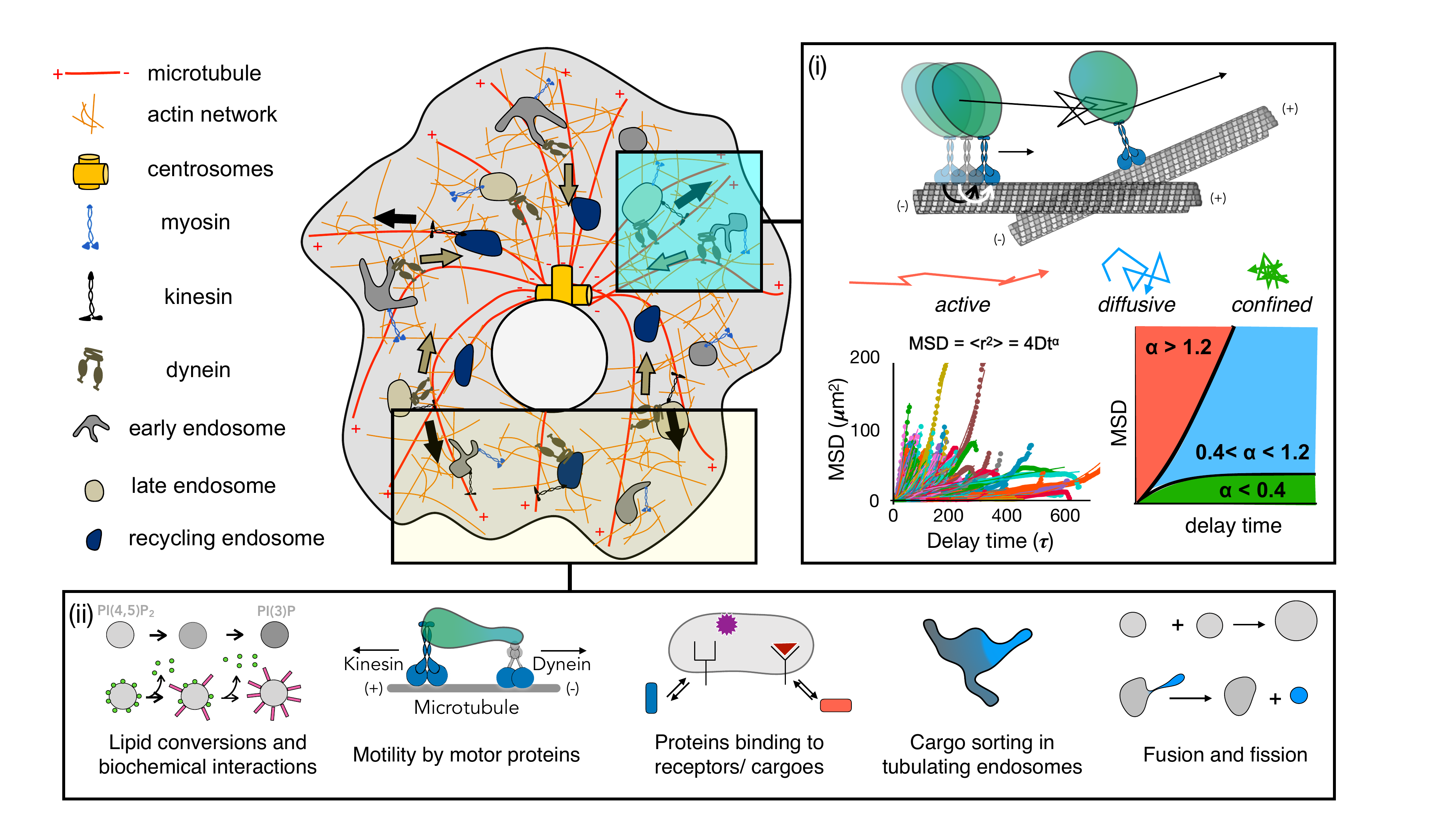}
  \caption{\textbf{Schematic of endosomal trafficking with key players including cytoskeletal elements, motor proteins, and endosomes.} (i) Motility may be characterized as active, diffusive, or confined. Note that mean squared displacement plots of Rab5 endosomes measured for five minutes display a distribution of motility characteristics, highlighting the stochastic nature of the endosomal system. (ii) Key processes include lipid conversions and biochemical interactions, motility by motor proteins, binding of proteins to receptors/cargoes, and cargo sorting in tubulating endosomes. Motility of any one kind of endosome within a time duration displays all kinds of motion, and one track displays multiple modes.}
  \label{fig:EndosomalTrafficking}
\end{figure}

Eukaryotic cells have a complex organization, which enables precise control of biochemical reactions through compartmentalization, both in membrane-bound organelles as well as through intracellular positioning. It is readily observable that cells show a heterogenous distribution of organelles and cytoskeleton, as well as cytoplasmic proteins and nucleic acids, which can be dynamically adjusted in response to cell identity and state. Cells utilize a vesicular transport system to move cargo between distinct intracellular locations including movements of cargo between organelles, of products destined to be released via exocytosis, and of materials internalized from the extracellular milieu to be delivered to specific destinations (Fig.~\ref{fig:EndosomalTrafficking}). Furthermore, these transport components have been increasingly identified to play critical roles in the intracellular positioning of almost every membrane-bound organelle.

Multiple layers of organization tightly regulate the transport and motility of these vesicles, ensuring the sorting and precise localization of internalized cargo. In addition to spatial movement, the identity of the transport vesicle, which is reflected in the lipid and protein composition of the cytoplasmic-facing membrane, shifts via endosomal conversion. This dynamic and stochastic process involves the exchange or conversion of proteins and/or lipids, leading to progressive biochemical maturation of the vesicle, and is crucial for the sorting and processing of cargo within the endosomal system~\cite{Cullen2008,RodriguezBoulan2005}.

\subsection{Cytoskeletal elements and motor proteins}\label{sec:activeTransportProteins}

The dynamic structure of mammalian cells is maintained through three complimentary cytoskeletal systems, composed of microtubules, actin, and intermediate filaments (IFs). Each of these filaments are composed of large polymerized proteins that self-assemble and disassemble in a dynamic fashion. They are able to anchor to membranes and transmit forces, providing structural rigidity and enabling cellular remodeling. By anchoring to organelles, the cytoskeleton also exerts control over subcellular localization, thereby playing a central role in cellular organization. While IFs are largely involved in nuclear structure and cell-cycle control, an emerging body of results implicates IFs in direct control of vesicular transport (reviewed by~\cite{margiotta2016}); furthermore, IFs are able to indirectly affect trafficking through crosstalk with the microtubule and actin cytoskeleton.

Long-range transport throughout the cell is primarily enabled by microtubules, elongated cytoskeletal filaments composed of polymerized tubulin dimers with inherent polarity, defined by a minus end at the microtubule organizing center (MTOC) near the nucleus, where the filaments are polymerized, and a plus end toward the plasma membrane, where tubulin subunits associate (and disassociate). Associating with these microtubules are the motor proteins dynein and the kinesin family, which bind to membrane vesicles as well as other organelles as part of multicomponent complexes. Following membrane binding, these motor proteins are able to ``walk'' along microtubules through sequential ATP-consuming cycles, which induce conformational changes that step the protein along the microtubule~\cite{Soldati2006}. These motor proteins sense the directionality of microtubules and step preferentially toward a specific end, with dynein showing minus end-directed trafficking and the kinesin protein family predominantly plus end-directed motility~\cite{Granger2014}.

\subsection{Membrane compartment identifiers}\label{sec:membraneCompartments}
A key component governing the structural and functional identities of endosomes is the family of membrane-bound proteins localized to the cytoplasmic-facing membranes. A notable example is the Rab family proteins. These are small GTPases that act as determinants of endosomal character enabling specific binding by a diverse range of proteins~\cite{Zerial2001}. The regulation of Rab GTP--GDP binding is crucial for controlling endosomal activity and involves various protein classes, including GDP exchange factors (GEFs), GTPase-activating proteins (GAPs), and GDP dissociation inhibitors (GDIs)~\cite{WandingerNess2014}. These recruited proteins, which bind to specific Rab proteins, influence the fate of the vesicle and its cargo by controlling features such as vesicular motility, via association with motor proteins and adaptors (Sec.~\ref{sec:activeTransportProteins}); membrane budding and tubulation, via membrane-shaping proteins (Sec.~\ref{sec:membraneSculptors}); and the modulation of vesicular fusion, via membrane-tethering proteins and the fusion machinery (Sec.~\ref{sec:effectorProteins}).

\subsection{Lipid composition}\label{sec:lipidComposition}

Endosomal identity is also determined by lipid composition, especially through key signaling lipids such as phosphoinositides (PIs). Despite being a small fraction of cellular membranes (less than 1\% of the total phospholipid pool), phosphoinositides play a critical role in organizing the membrane structure of the vesicular transport system, in addition to the plasma membrane, Golgi apparatus, and endoplasmic reticulum (ER)~\cite{Posor2022,Wallroth2018}. These membrane phospholipids consist of a myo-inositol ring that can undergo reversible phosphorylation and dephosphorylation at the 3-, 4- and 5-OH groups via specific kinases and phosphatases. These enzymes are recruited to their target organelles through association with specific membrane-bound proteins such as Rab proteins. This establishes a spatial heterogeneity of phosphoinositides across cellular sub-compartments, which become associated with specific vesicular populations within the cell~\cite{Jovic2014}. The interconversion of these lipids is highly dynamic and changes as vesicles mature and lipids are exchanged, such as through membrane fusion~\cite{Cullen2014}. As such, these lipids are able to act as molecular ``signposts'' to orchestrate the spatiotemporal recruitment of membrane proteins containing a domain that recognizes a particular phosphoinositde (e.g., the PH and FYVE domains). Despite their essential role in cellular organization and the growing list of diseases associated with phosphoinositide dysfunction, the mechanisms by which phosphoinositide conversion is spatiotemporally controlled remain largely unknown. These aspects are reviewed in~\cite{Wallroth2018,Posor2022}.

\subsection{Effector proteins}\label{sec:effectorProteins}
A diverse array of endosomal effector proteins interacts with Rab proteins, phosphoinositides, and the cytoplasmic face of transmembrane receptors located within the endosomal membrane. These effector proteins typically exhibit weak affinity for endosomes, allowing for competition and exchange among effectors that recognize the same sequences. Stable binding of effectors can be influenced by factors such as clustering, recognition of membrane curvature, and coincidence detection of specific cognate protein and phosphoinositide species on the same membrane~\cite{Carlton2005}. Once bound, these effector proteins can influence cargo transport through endosomal tethering and fusion, as well as cargo sorting followed by subsequent fission.

\subsection{Membrane sculpting, fusion, and fission}\label{sec:membraneSculptors}
Alongside vesicular motility, cargo transport is dependent on transit through correct endosomal compartments, with most cargoes typically passing through multiple endosomal populations en route to their destination. A major step in this process is sorting of cargoes at the early endosomal level by tubulation and fission, as well as newly generated compartments fusing into the next set of compartments. Membrane deformation leading to tubulation, reshaping, or scission results from the action of motor proteins that pull on membranes~\cite{Campas2008}, actin polymerization on the membrane by actin nucleators~\cite{Derivery2009}, or the interaction of curvature-inducing membrane-binding proteins, collectively termed here as membrane sculptors~\cite{Mim2012}. Endosomal fusion is dependent on Rab GTPases, tether molecules such as EEA1 and SNAREs. The main superfamily of such proteins is the Bin/amphiphysin/Rvs (BAR) domain containing family, which includes proteins that control membrane curvature in endosomal fission, maturation, and endocytosis; these proteins are also found in other organelles such as mitochondria and at the plasma membrane~\cite{Simunovic2015}. Other organelles also play a role in endosomal fission: cargo-containing tubule is scissioned from the parent endosome through the action of the endoplasmic reticulum and the actin cytoskeleton~\mbox{\cite{Derivery2010,Kar2023,Rowland2014,Striepen2022}}.

\section{STOCHASTIC MODELS FOR INTRACELLULAR TRANSPORT}\label{sec:stochasticModels}

The complex interplay between membrane-bound proteins, phosphoinositides, and effector proteins---as well as enzymes, motor proteins, force-generating proteins, and cargo-bound transmembrane proteins---enables the precise transport of materials within the cell, regulates vesicular--organellar interactions (such as lysosomal fusion), and modulates interactions with cytosolic proteins, in addition to signal processing of receptors that are internalized following activation. Yet, each of these processes has intrinsic stochasticity. (Basic aspects are reviewed in the Appendix A.) The focus of this review is how deterministic outcomes emerge in whole-cell phenomenologies despite noisy constituent dynamics.  In addition to the stochastic models discussed in this review, the mean completion times for many other transport processes have also been modeled, such as the mean search times of particles diffusing on a network with given properties~\cite{brown2020}. However, we restrict ourselves here to examples of processes in which either the full distribution or the fluctuations about the mean of the relevant variables have been characterized, and refer the reader to the excellent review by Mogre et al. for details on other models~\cite{Mogre2020}.

\subsection{Undirected transport processes}\label{sec:undirectedTransport}
\begin{figure}[h]
   \includegraphics[width=4.5in,page=2,trim={5in 1.3in 5in .4in},clip]{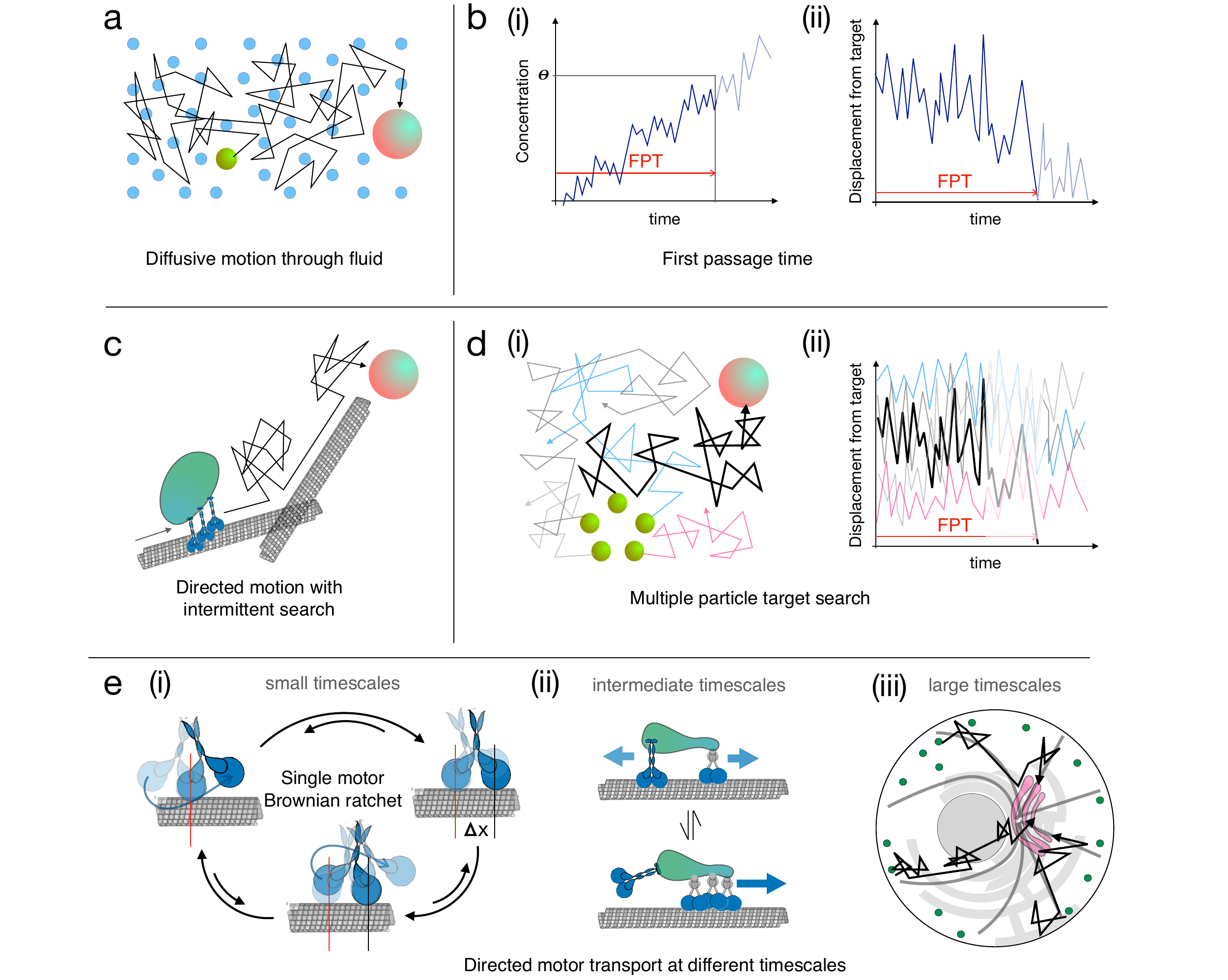}
    \caption{\textbf{Schematic of intracellular transport processes and characterization of timing of events.} \textbf{(a)} A particle (green) undergoes diffusive motion in the cytoplasmic fluid due to random collisions with fluid particles (blue), until it reaches its target (big sphere). \textbf{(b)} The timing of events in the cell can be characterized by their first passage time (FPT), such as the FPT of \textbf{(i)} the concentration of a certain species to reach a threshold ($\theta$) and \textbf{(ii)} a moving particle to reach a target, which is similar to (a). \textbf{(c)} Directed motion with intermittent search may involve a cargo (green) undergoing alternating periods of directed motion with the help of motors (blue) along a 1D microtubule (gray) and diffusive motion when detached from the microtubule, until it reaches its target. \textbf{(c)} In the case of multiple particles searching for the same target, the FPT of the target being found is the minimum of the FPTs of the individual particles. \textbf{d} The directed motion with intermittent search in (b) behaves differently at different timescales. \textbf{(i)} At the smallest timescale, the motion of a single motor along a microtubule takes place through an energetically biased cycle of changing conformal states, modeled by the Brownian ratchet. \textbf{(ii)} At intermediate timescales of a few cycles, the motion of a single motor can be well-approximated by 1D advective motion with constant diffusion and drift. The state of motion of the cargo, which has multiple motors attached to it, is determined by which of those motors are attached to the microtubule. The individual motors stochastically attach to and detach from the microtubule, leading to a change of state, an example of which is shown. \textbf{(iii)} At large timescales, the cargo performs directed motion with intermittent search, as shown in (b).}
    \label{fig:models}
  \end{figure}

Biological systems utilize various modes of transport, which we broadly classify as ``undirected'' and ``directed''. These processes typically possess different effective descriptions characterizing motion over different length scales, ranging from nanometers to micrometers (Fig.~\ref{fig:models}).

\subsubsection{Diffusion (passive)}\label{sec:diffusion}
At the shortest scales of motion, all biomolecules undergo diffusion, a passive mode of transport driven by conversion of fast thermal fluctuations of light molecules, such as water, in the surrounding medium into comparatively slower motion of the heavy biomolecule under observation~\cite{1905-einstein-brownian,Bouchaud1990}. Heavy biomolecules moving through the cytoplasmic medium continually collide with lighter fluid molecules, leading to a disjointed trajectory termed Brownian motion, characterized by random small ballistic (inertial) movements between successive collision events (Fig.~\ref{fig:models}a). Over timescales slower than those characterizing these rapid collisions, a simpler behavior emerges, which can be modeled using two complementary perspectives~\cite{Bressloff2013,Gardiner2009}: the Langevin equation governing the random motion of the biomolecule position $x(t)$,
\begin{equation}\label{eq:diffusion}
    \dot{\vb{x}}(t)=\frac{\vb{F}(\vb{x}(t))}{\gamma}+\pmb{\eta}(t),
\end{equation}
and the equivalent Fokker-Planck (FP) equation governing the time evolution of the probability density of locating the biomolecule:
\begin{equation}\label{eq:diffusionFP}
    \partial_tp(\vb{x},t)=\div\left[\left(-\vb{F}(\vb{x},t)/\gamma + D\grad\right)p(\vb{x},t)\right].
\end{equation}
Herein, $\vb{F}(\vb{x})$ is a generic external force field; $\gamma$ is the effective inverse mobility of the biomolecule in the medium, yielding a drift velocity $\vb{F}(\vb{x}(t))/\gamma$; $D$ is the diffusion coefficient or diffusivity; and each Cartesian component of $\pmb{\eta}(t)$ characterizes an independent Gaussian white noise with variance proportional to $2D$: $\le\la \eta_{i}(t) \eta_{j}(t') \ri\ra = 2D \delta(t-t')$. The mobility and diffusivity are related through the Einstein relation, $\gamma D = k_{B}T$~\cite{1905-einstein-brownian}, where $k_{B}$ is Boltzmann's constant and $T$ is the temperature of the medium. Since $\gamma$ varies linearly as the size of the biomolecule (generalized Stokes' law), $D$ is a decreasing function of molecular size.

When the external force is zero, the motion of the biomolecule is characterized as (undirected) diffusion. A characteristic property of undirected diffusive motion is that the mean square of each (Cartesian) component of displacement of a single biomolecule from its starting location increases linearly with time~\cite{1905-einstein-brownian}:
\begin{equation}\label{eq:MSDdiffusion}
    \le\la (x_{i}(t) - x_{i}(0))^{2}\ri\ra=2 D t.
\end{equation}
One can contrast this with simple ballistic motion, where the mean square displacement grows as the square of elapsed time. When large numbers of the same biomolecule undergo diffusion, $p(\vb{x},t)$ can be replaced by the number density of biomolecules, $n(\vb{x},t)$, in Eq.~\eqref{eq:diffusionFP}, yielding the familiar diffusion equation (Fick's second law) governing the passive spreading of biomolecules in a fluid medium:
\begin{equation}\label{eq:diffusion0}
    \partial_t n(\vb{x},t)=\div\left[\left(-\vb{F}(\vb{x})/\gamma + D\grad\right)n(\vb{x},t)\right].
\end{equation}

Since diffusion speeds up for smaller biomolecules, as discussed above, small proteins of 1--5 nm such as motor and adaptor proteins diffuse freely and efficiently, but vesicles (100--1000 nm) are largely confined, with their free motion further compromised by molecular crowding~\cite{Bonucci2023,LubyPhelps2013,Seksek1997}. However, diffusion still plays a crucial role in vesicular transport; for example, in motor proteins dispersed in the cytoplasm searching for binding partners and microtubules, and vesicles themselves searching for targets not located directly on the microtubules (see Sec.~\ref{sec:intermittentSearch}).

Diffusion is not as useful as energy-consuming ballistic motion for long range movement of cargo by large objects such as vesicles, but the situation is different for smaller molecules. Even accounting for the reduced diffusion of molecules within the crowded and active intracellular environment, the average time taken for a small protein to diffuse from the plasma membrane to the perinuclear region within a cell ($\sim 10\mu$m) is on the order (or faster) of the time taken for a motor protein to cross this distance (moving at $\sim 800$ nm/s \textit{in vivo})~\cite{presley1997}.  Purely considering the speeds of these processes, it appears counterproductive to encapsulate cargo into larger, constrained vesicles. However, the control of interactions through compartmentalization and directed transport is essential to cellular organization, which limits spurious interactions and mislocalization of cargo. For example, signals are more faithfully transmitted from activated receptors to the nucleus by trafficking groups of endocytosed receptors toward the perinuclear region than by activating secondary messengers in the periphery of the cell, which must diffuse toward the nucleus and thus participate in additional interactions such as dephosphorylation and deactivation~\cite{Villasenor2016}. The simple diffusion process is also popular as a theoretical model since intuition-enhancing analytic solutions exist for many problems; see Appendix A for extended discussion.

\subsubsection{{Subdiffusion}}\label{sec:subdiffusion}
For many intracellular transport processes where biomolecules passively move through a complex quasi-fluid medium, the mean square displacement grows sublinearly with time,
\begin{equation}\label{eq:MSDsubdiffusion}
    \le\la (x_{i}(t) - x_{i}(0))^{2}\ri\ra\propto t^{\alpha}, \quad 0<\alpha<1.
\end{equation}
This kind of motion, termed subdiffusion, lies between the limits of simple diffusion ($\alpha = 1$, see Eq.~\eqref{eq:MSDdiffusion}) and ``caged'' motion ($\alpha = 0$, corresponding to the biomolecule ``rattling'' inside a small bounded region). As expected, subdiffusive motion is much slower than ballistic motion, which is characterized by $\alpha=2$.

Subdiffusive motion is found within the crowded cytosolic environment, where soluble proteins, lipids, cytoskeletal filaments, and organelles occupy up to 50\% of the volume~\cite{Dix2008}. It is extremely inefficient as a mechanism for long-range transport of material. Examples include the motion of messenger RNA molecules~\cite{golding2006}, chromosomal loci in bacteria~\cite{weber2010}, and lipid granules in yeast cells~\cite{tolic2004,jeon2011}. A simple rule of thumb for passive motion inside the cell is that smaller particles (corresponding to tens of nanometers in size) tend to exhibit diffusive motion, while particles an order of magnitude larger tend to exhibit subdiffusive motion~\cite{Mogre2020}.

Two mechanisms are commonly invoked to explain the emergence of subdiffusion~\cite{Bressloff2013}. The first theory assumes that biomolecules pause briefly at binding sites between patches of diffusive motion, giving rise to a non-exponential distribution of wait times between steps in the discrete Brownian motion framework. Such processes are typically modeled as continuous time random walks~\cite{scher1975} and are weakly non-ergodic. In the second theory, the cytoplasm is assumed to behave like a viscoelastic fluid due to the presence of elastic elements such as nucleic acids and cytoskeletal filaments. This structure leads to long-term correlations in noise, like a form of medium memory~\cite{2019-nagel-RMP}. The corresponding motion of the biomolecule can then be modeled using fractional Brownian motion or a fractional Langevin equation~\cite{Burov2011,Mandelbrot1968}. 

In addition to these mechanisms, diffusive motion in the presence of a high concentration of obstacles can also appear subdiffusive at short timescales~\cite{Mogre2020}. Theories of (sub)diffusion can also include molecular crowding, traps, and confinement~\cite{Bressloff2013}. Some studies suggest that multiple mechanisms can coexist~\cite{jeon2011,weigel2011}. Thus, determining which mechanism is the source of the experimentally observed subdiffusive behavior cannot be accomplished by simply considering the scaling of variance. Other measures, such as ergodicity, are needed to deduce the mechanisms underlying a specific instantiation of subdiffusion.

\subsubsection{Active diffusion}\label{sec:activediffusion}

The densely packed cytoplasm and cytoskeletal filaments enable an alternative mode of transport, termed active diffusion. This random intracellular motion is achieved by the active force fluctuations of these cytoskeletal elements~\cite{Drechsler2017}. The summation of many ATP-driven contractile processes leads to random motion of particles, albeit at higher displacements than those of constrained diffusion due to the ``stirring'' of the cytoplasm~\cite{Guo2014}. Active diffusion is important in a range of systems, especially in larger cells with sizes up to 100 \textmu m. Active diffusion can be much faster than passive diffusion due to an enhancement of the diffusion constant (the value of $\alpha$ still remains $1$ when excluding effects that lead to subdiffusion or drift), and can additionally be modulated in different cells, or even spatially within a single cell~\cite{Brangwynne2009}.

\subsection{{Directed transport}}\label{sec:directedTransport}
Although diffusion is effective for transport over short distances without incurring energetic costs, it becomes inefficient over longer intracellular distances for large objects such as vesicles. Additionally, its unbiased nature precludes spatial sorting. These challenges are overcome via the directed transport processes detailed below. Through the active expenditure of energy, these directed transport processes permit precise and directed movement of vesicles and cargo~\cite{2009-hirokawa,2013-roberts}.

\subsubsection{Advection}\label{sec:advection}

Advection refers to a net overall flow of the cytoplasm in a particular direction, usually due to actomyosin cortex-generated flows. In the simplest models, the cytoplasm is treated as a linearly viscous (i.e., Newtonian) fluid and the advective effect of net flow is incorporated into a background drift velocity, $\vb{v}_{\text{fl}}$, which replaces the $\vb{F}/\gamma$ term in Eq.~\eqref{eq:diffusionFP}. The inclusion of this effect makes the mean square displacement increase quadratically with time at large times (i.e., $\alpha=2$). However, the fluctuations about the time-dependent mean continue to be diffusive, i.e., $\le\la (x(t) - \le\la x(t)\ri\ra)^{2}\ri\ra \propto t$, which dominates the mean square displacement at short times. The relative importance of drift over diffusion for a particle's advective motion over a lengthscale $L$ is given by the dimensionless P{\'e}clet number, $\text{Pe} = L|\vb{v}_{\text{fl}}|/D$. Thus, the drift term dominates at longer lengthscales ($\text{Pe}\gg 1$), while the motion is predominantly diffusive at shorter lengthscales ($\text{Pe} \ll 1$). Examples where motion is predominantly dominated by drift due to long lengthscales include cytoplasmic streaming in plant cells~\cite{tominaga2013,woodhouse2013} and \textit{Drosophila} oocytes~\cite{ganguly2012}.

The preceding model can be improved by treating the cytoplasm as a poroelastic material consisting of a fluid phase interacting with an elastic solid phase~\cite{Mitchison2008,Mogilner2018}. The upgraded treatment has been shown to more accurately reproduce the flow patterns arising due to blebbing, motility, indentation, and cytoskeletal contractions~\cite{Charras2008,Lewis2015,Moeendarbary2013,Radszuweit2013}.

\subsubsection{Motor protein driven cargo movement}\label{sec:motorTransport}

The directed motion of cargo along cytoskeletal filaments such as microtubules requires energy expenditure in successive ATP-consuming cycles, which power specialized motor proteins capable of stepping along the filaments. Active motor transport has different effective descriptions at short, intermediate and long length- and timescales as discussed below~\cite{Bressloff2013,Julicher1997,Keller2000,Klumpp2005,Kolomeisky2007}.

\paragraph{Short time scales: Stochastic motor movement}

At the microscopic level, a stochastic motion of the motor protein on a filament, driven by thermal fluctuations and energy (ATP) consumption, can be modeled by the Brownian ratchet mechanism~\cite{Reimann2002}. In this model, the motor stochastically jumps between different conformational states while moving on a single track (cytoskeletal filament) labeled by a one-dimensional (1D) coordinate $x$. In a given conformal state $i$, the motor undergoes Brownian motion inside a periodic potential $V_{i}(x)$ with diffusion constant $D_{i}$. Both these quantities are specific to a given state, but all potentials have the same period, which is equal to the step size of the motor $L$. The central idea is that the states are ordered in such a way that the minima of the potentials are successively shifted forward. Thus, when transitioning cyclically between the states in that same order, after each transition rests long enough to slide into the forward-shifted minimum of the new potential, the motor can move forward step by step. A quasi-realistic three-state model of such a motor is sketched in Fig.~\ref{fig:models}e (i).

When no energy is supplied, detailed balance in thermodynamic equilibrium ensures that there cannot be any sustained directed motion of the motor, irrespective of the construction of its different conformal states (see Appendix A). However, motor proteins consume energy using ATP molecules in a nonequilibrium process that allows them to move in a directional manner along cytoskeletal filaments~\cite{Bressloff2013,Julicher1997,Liepelt2007,Lipowsky2005,Parmeggiani1999}. Incorporating the overall motor velocity into the description of motor motion over longer timescales ranging over multiple steps, motor motion can be well-approximated by simple diffusion with drift~\cite{Peskin1995,Wang1998}, represented through Eq.~\ref{eq:diffusion}. On this timescale, detailed information about conformal states is averaged out.

\paragraph{Intermediate time scales: Cargo attached to multiple motors}\label{sec:multipleMotorTransport}

As vesicular cargo moves along an intracellular cytoskeletal filament, it is usually attached to and moved by multiple motors of different types (Fig.~\ref{fig:EndosomalTrafficking}). Although the motion of each individual motor is biased toward a particular direction, this direction could vary between different motors. For example, kinesin moves towards the $(+)$ end of the microtubule, while dyenin moves towards the $(-)$ end~\cite{howard2001} (Fig.~\ref{fig:EndosomalTrafficking}(ii)). Moreover, motors stochastically attach to and detach from the filament and cargo, allowing the motion of cargo to vary in both magnitude and direction. At any given time, the configuration of the attached motors determines the net diffusivity and drift velocity of the cargo (Fig.~\ref{fig:models}e (ii)). Depending on the specifics of the model, different kinds of realistic cargo motion have been predicted. 

Using the fact that motor proteins can work cooperatively or antagonistically, a phenomenon known as ``tug-of-war'', characterized by bidirectional transport and stochastic stalling, has been proposed~\cite{Schnitzer2000}. In the tug-of-war model, where kinesin and dynein motors exert opposing forces, the drift velocity of a particular state in the motor--cargo system is approximately linearly dependent on the net force resulting from the specific motors attached to the filament in that state~\cite{Visscher1999}. However, \textit{in vivo} observations suggest that dynein and kinesin may also exhibit inhibitory protein-protein interactions that contribute to stalling behavior, necessitating a more complex model than purely mechanical opposition~\cite{Gennerich2009}. An extension of this model that accounts for interactions with other particles moving along the same track, via the specific constraint that particles cannot occupy the same site simultaneously, is the totally asymmetric exclusion process (TASEP). The simplest version of this model allows for finding exact solutions for the stationary state~\cite{Blythe2007,Chou2011,Schadschneider2010}. Additional factors such as absorption and desorption kinetics can be incorporated but require mean field approximations or numerical methods to solve.

\paragraph{Long timescales: Directed motion with intermittent search}\label{sec:intermittentSearch}

Over long timescales when transit lengths approach the lengthscales separating source and target for cargo motion, the linear movements of motor protein-driven vesicles along microtubules are punctuated by frequent pauses, including periodic ``turns'' as motors desorb from one microtubule and hop to a nearby microtubule. This process, characterized by frequent pauses and intermittent bidirectional motility, is termed a ``random intermittent search process'' (Fig.~\ref{fig:models}c). Such motion mimics efficient search strategies observed in animal foraging~\cite{Bartumeus2005}. The frequent pauses observed in motor protein-driven vesicles may enable exploratory forays into the cytoplasm to locate target destinations~\cite{Benichou2011}. Since most intracellular targets are not located at the MTOC, vesicles need to detach from microtubules in the vicinity of the target organelle anyway. Although the reaction kinetics during processive motion may not favor microtubule desorption, the diffusive state allows for a better match between vesicle residence time and lower-affinity binding, enabling a more accurate interpretation of local reaction space by ensuring the required reaction kinetics can take place. Thus, by incorporating stochastic unbinding, the entire intracellular space can be explored, enabling ``searching'' for the target molecules, which can be both cytoplasmic or membrane-bound, occurring through organelle-organelle interactions, localized within a particular sub-cellular localization. Prominent examples include phosphatases that are required to attenuate receptor signaling, and fusion with lysosomes that are predominately found within the perinuclear region of the cell~\cite{Stanoev2018}.

At these long time- and length-scales of vesicular motion where ``searching'' is necessary, the biologically relevant quantity is the First Passage Time (FPT) distribution of the search process~\cite{Gardiner2009,2016-iyer-biswas-FPT,2014-iyer-biswas-PRL,2017-iyer-biswas-intthresh} (Fig.~\ref{fig:models}b). For a single searcher, this usually yields results similar to that obtained for the 1D diffusion process (see Appendix A). As discussed in Sec.~\ref{sec:FPTn} below, the search process becomes faster and more deterministic when multiple searchers are involved. This strategy is utilized when multiple cargoes are destined for the same target.

\subsection{Vesicular maturation and sorting}\label{sec:maturation}

Vesicular maturation is defined as an endosome losing a specific molecular identifier and gaining a new one; for example, APPL1 to EEA1~\cite{Zoncu2009} or Rab5 to Rab7~\cite{Rink2005}. Although these studies uphold a single endosome-centric view of maturation, it is evident that fission and fusion processes continually occur. Furthermore, the tubulation leading to fission may also involver a cargo sorting step that provides an additional functionality to move the cargoes in synchrony with the maturation. 

Vagne and Sens~\cite{Vagne2018a} have modeled vesicular transport and cisternal maturation through a sequence of irreversible steps in which (a) a membrane-bound compartment receives an influx of a given component A through homotypic fusion (i.e., fusion that occurs when A is already present on the vesicle); (b) A subsequently converts to B through the maturation process; and finally (c) B exits through selective budding. For examples of such processes, see \cite{Binder2012,Castro2021,Foret2012,Zeigerer2012}. Modeling these processes as a sequence of elementary Markovian reactions, under the extremely simplified assumption of constant rates, the mean FPT can be determined analytically~\cite{Vagne2018a} while the full stochastic distribution is found numerically via the standard Gillespie algorithm~\cite{Gillespie1976}. The steady-state dynamics were found to be controlled by two parameters: $r_{1}$, the ratio of the rate of vesicle injection to that of budding, and $r_{2}$, the ratio of the rate of conversion (from A to B) on the compartment's surface to that of budding; $r_{1}$ controls the size and $r_{2}$ the composition of the vesicle~\cite{Vagne2018a}.

An implicit assumption in the above model is that the components A and B do not tend to cluster on the endosomal surface (i.e., they are not more clustered than a random distribution). Recent experiments investigating the early endosomal maturation characterized by the conversion from APPL1 to EEA1 effector proteins have shown that these proteins form homotypic clusters on the surface of the endosome instead of attaching at random locations~\cite{York2022}. To account for a possible utility of such homotypic clustering, agent-based simulations of the maturation process were performed by enhancing the above simple model with processes that prefer homotypic clusters on the endosomal surface. Through the use of agent-based simulations, the effects of clustering and collisions on the timing of the early endosomal maturation process has been quantified in~\cite{York2022}, with the simulation results suggesting that this clustering mechanism significantly reduces the mean and variability in the conversion time (for further details see Secs.~\ref{sec:clustering} (clustering) and~\ref{sec:noiseBenefits}).

\section{MITIGATING VERSUS UTILIZING NOISE}

It is natural to view noise as detrimental to order. However, biological systems do not merely have to overcome the stochasticity inherent in molecular interactions---they sometimes also gain advantage from it~\cite{Wehrens2018}. The intracellular endosomal trafficking network presents a clear example of a system with significant levels and diverse sources of stochasticity, yet an overall robustness in reliability of cargo transport; for example, delivery of cargoes to lysosomes for degradation~\cite{Cullen2018} or from the endosomes to the Golgi network~\cite{Bonifacino2006}.

\subsection{Noise suppression}\label{sec:noiseSuppression}

\subsubsection{Mitigation of noise through use of large copy numbers}

\paragraph{Large numbers of chemical species: Pooling}\label{sec:pooling}

A widely prevalent strategy for reduction in noise follows from the law of large numbers (see Appendix A). Consequently, when large numbers of biochemicals are present in a cell, their transport or reactive behaviors can be deterministic even though each individual microscopic step is stochastic. Thus, even though a single diffusing biomolecule executes an irregular path governed by a single instantiation of the time evolution of the Langevin equation, Eq.~\eqref{eq:diffusion}, when a large number of such molecules are considered, their density follows the deterministic diffusion law, Eq.~\eqref{eq:diffusion0}. The noise in the motion of individual particles is perceived as being eliminated. Similar arguments can also be put forward for chemical reactions between large numbers of biomolecules, when the reactant numbers evolve in a deterministic fashion despite the inherent, often large, stochasticity present in biochemical processes at the molecular level~\cite{2009-iyer-biswas,2014-iyer-biswas-mixedP,2009-iyer-biswas-dissertation,2009-iyer-biswas-powerlaws}. Thus, when reactants are present in large numbers, cellular processes that depend on them (yet are composed of ubiquitous molecule-level steps of chemical reactions and stochastic transport) become deterministic.

A modified strategy of ``pooling'' can also be used to lower noise in biochemical processes where some critical components are present in small numbers, such as during genetic transcription and translation~\cite{2009-iyer-biswas,2014-iyer-biswas-mixedP,2009-iyer-biswas-dissertation}. The noise originating from precursor processes can be suppressed by maintaining a reservoir (pool) of necessary substances from those processes. The maintenance of noise-free deterministic precursor chemical processes, such as one in chemical equilibrium, ensures a steady (constant in the case of chemical equilibrium) supply of the precursors, allowing that chemical noise to be absent in subsequent processes. For example, York et al.~\cite{York2021} showed that there exists an epidermal growth factor (EGF)--Ca$^{2+}$--APPL1 interaction that leads to the rapid desorption of APPL1 from pre-existing endosomes and the binding of re-binding APPL1 via a distinct phosphotyrosine binding domain to freshly generated endosomes containing phosphorylated EGF receptor (EGFR). This then allows dynein recruitment and the highly processive re-localization of these endosomes to the ER-rich perinuclear region, which has been shown to facilitate EGFR deactivation~\cite{Stanoev2018}. This EGF--Ca$^{2+}$--APPL1--dynein nexus thereby leads to the tight control of the EGFR signaling window in response to large concentrations of EGF, imparting robustness to the cell's growth factor sensing.

\paragraph{Large numbers of searchers}\label{sec:FPTn}

Search processes can also utilize the presence of a large number of searchers for faster and more deterministic detection of a target (when compared to the same search being performed by a single searcher). Search processes can be modeled as FPT processes (Fig.~\ref{fig:models}d), with the FPT for a single particle obtained from a stochastic process using the formalisms discussed in Sec.~\ref{sec:firstPassageTime}. Given the FPT $P(\tau)$ for a single particle, the FPT of $N$ independent searchers becomes~\cite{Bressloff2012}:
\begin{equation}\label{eq:FPTn}
    P^{(N)}(\tau)=N P(\tau)\left[1-\int_0^\tau P(\tau')d\tau'\right]^{N-1}.
\end{equation}
Interestingly, this expression arises in a different biological context---that of emergent periodicity in synchronized flashing of fireflies~\cite{Sarfati2023}, where both the mean and the variance of the net FPT distribution, $P^{(N)}(\tau)$, decreases as $N$ increases, irrespective of the specific model underlying $P(\tau)$. Thus, increasing the number of searchers makes the search process faster and more deterministic. For more detailed reading, such as calculation of the asymptotic limits of the composite FPT distribution of a large number of searchers starting from known single-particle FPT distributions as in Eq.~\eqref{eq:FPT1Ddiffusion}, see~\cite{Bressloff2012,Bressloff2013}.

\subsubsection{Spatiotemporal organization strategies}

\paragraph{Clustering}\label{sec:clustering}

Clustering of molecules on the surface of vesicles also plays an important role in aiding the directed motion of the vesicles themselves through the cytoplasm. For example, it has been shown that the clustering of dynein motors on the membrane of a phagosome allows for the generation of a cooperative force on a single microtubule, resulting in rapid directed transport of the phagosome along the microtubule~\cite{rai2016}, overcoming the stochasticity associated with opposing motor proteins on an endosome that results in bidirectional motility. The effect of multiple motors attached to a cargo is modeled in Sec.~\ref{sec:multipleMotorTransport}.

Clustering can lead to specificity and increases recruitment rates, as exemplified by recruitment of dynamin through the clustering of phosphoinositides~\cite{picas2014}. Although the effect on conversion time in the absence of clustering has not been experimentally measured (due to the lack of methods to selectively turn off the self-affinities of the proteins under consideration), simulations  show  that  turning  off  clustering  while  keeping all other rates constant significantly increases the mean and variance of the conversion time~\cite{York2022}. The clustering of phosphoinositides has also been postulated to be involved in experimentally measured EEA1 clustering~\cite{York2022}, that plays a role in endosomal conversions. In a new proposed model of seeded endosomal conversions, incoming APPL1 endosomes collide with pre-existing mature EEA1 endosomes, which results in ``transfer'' of EEA1. Clustered EEA1 on mature endosomes ensures threshold number of molecules being planted onto the incoming endosome. Agent-based simulations show that  clustering significantly reduces the mean and variance in the conversion time in endosomal maturations~\cite{York2022}.

\paragraph{Hierarchical arrangement of timescales}\label{sec:separationTimescales}

The separation of timescales along with the energetic coupling between different molecule types being transported plays an important role in suppressing noise in membrane transport mechanisms, which involve the transport of molecules between two compartments separated by a membrane. Such mechanisms have been used to model the suppression of noise in intracellular glucose levels through sodium-potassium pumps in combination with sodium-glucose coupled transporters~\cite{Cardelli2020}. In short, the reaction noise of the transported molecule of interest, say `$A$', on the `target' side (II) of the membrane is reduced when the timescale (lowering the rates) of transport across the membrane is increased to well above the timescales of reaction noise in $A$ on the `source' side (I) of the membrane. 

We discuss below simple examples of this phenomenon when the molecule $A$ undergoes a bursty birth-death reaction on the source side:
\begin{align}\label{rxn:uniporterProduction}
    \ce{$\phi$ ->[b_A]v_A A_{\text{I}}};\quad \ce{A_{\text{I}} ->[d_A] $\phi$},
\end{align}
where the subscript on $A$ denotes the side of the membrane the molecule is located in. In steady state,  this process has a super-Poisson Fano factor $F=(1+v_A)/2$ (see Appendix A). The dynamical timescale is controlled by the death rate, $d_{A}$~\cite{2009-iyer-biswas}.

\textbf{Uniporter dynamics: } A simple uniporter reversibly transports a species $A$ between compartment I and II as follows~\cite{Cardelli2020}: $\ce{A_{\text{I}} <=>[k][r \times k] A_{\text{II}}}$.
The value of $r$ controls the equilibration ratio for transport and $k$ sets its inverse timescale. The Fano factor on side II is~\cite{Cardelli2020},
\begin{equation}
    F_{A_{\text{II}}}=1+\frac{(v_A-1)}{2(1+r +d_A/k)}.
\end{equation}
Clearly, noise in $A$ is suppressed on side II (the Fano factor decreased) while holding equilibrium concentrations of $A$ constant, if $k$ is made much smaller than $d_{A}$, i.e., if transport occurs slowly compared to $A_{I}$'s fluctuation dynamics.

\textbf{Adding nonlinear coupling. Symporter+Antiporter dynamics: }  The uniporter is unable to achieve sub-Poisson level noise suppression (when $ F_{A_{\text{II}}}<1$). Symporters and antiporters can exceed this limit of performance by the coupling the transport of $A$ with that of another molecule, $B$~\cite{Cardelli2020}:
\begin{subequations}
    \begin{align}
        &\text{Symporter:}& &\ce{A_{\text{I}} + B_{\text{I}} <=>[k][r\times k] A_{\text{II}} + B_{\text{II}}},\\
        &\text{Antiporter:}& &\ce{A_{\text{I}} + B_{\text{II}} <=>[k][r\times k] A_{\text{II}} + B_{\text{I}}}.
    \end{align}
\end{subequations}
Assuming that $B_{\text{I}}$ obeys chemical dynamics similar to $A_{\text{I}}$ (Eq.~\ref{rxn:uniporterProduction} with rates with subscript $B$), for the simple case when $v_A=v_B=v$, as $k$ varies from 0 to $\infty$, the Fano factor of $A_{II}$ varies between:
\begin{equation}
    \frac{1}{2}\leq F_{A_{\text{II}}}\leq\frac{(1+v)(b_A+b_B)+4\sqrt{b_Ab_Br}}{4(b_A+b_B+2\sqrt{b_Ab_Br})}.
\end{equation}
Thus, by combining a slow transfer process ($k \ll d_{A}, d_{B}$) with appropriate coupling between A and B, symporters and antiporters can suppress the noise in $A$ to as low as sub-Poisson levels on side II of the membrane!

\paragraph{Coincidence detection}\label{sec:coincidenceDetection}

Coincidence detection describes the weak binding of a protein to two or more nodes (such as proteins and phosphoinositides). A variety of proteins that are specific to endosomal surfaces bind via coincidence detection of binding partners and specific phosphoinositide-binding via specialized motifs such as the PH Bar domain or FYVE domains~\cite{Carlton2005}. Thus, such proteins only effectively localize to membranes that contain all binding partners, thereby increasing specificity by reducing spurious binding. This also decreases the number of components required to facilitate effector localization, due to a sharp increase in the number of permutations available to enable the effective and coordinated binding and unbinding of proteins.

\paragraph{Structural organization}

Specific structural features of biomolecules can enhance their affinity to other biomolecules, reducing noise in associated processes. For example, in the case of proteins diffusing across a thermally fluctuating membrane, mismatch between the curvature preferred by the proteins and the surrounding membrane curvature can guide and modulate their motion, imparting greater precision and enhancing lateral diffusion~\cite{Brandizzi2002}. In the cisternal maturation model (cf. Sec.~\ref{sec:maturation}), the affinity-driven process of homotypic fusion relies on specific interactions between identical components, promoting the fusion of membranes and facilitating maturation. Nanoclusters of activated receptors (phosphorylated EGF receptors) also form discrete packages of signaling information that provide a robust signal via ``analogue-to-digital conversion''~\cite{Villasenor2015}.

\subsubsection{Reaction cascades}\label{sec:reactionCascades}
Endosomal networks transmit and decode extracellular signaling events and so must be specific, multiplexed, robust, and adaptable~\cite{York2020}. The network must translate a given input into a specific output, simultaneously process different signals, and resist both internal and external fluctuations---all while tuning itself to suit the cellular identity and state. Propagation of noise between interdependent processes depends on network circuit topology and timescales~\cite{Bruggeman2018,Villasenor2016}. Many theoretical studies have examined features of network topology that enable signaling cascades to confer emergent behavior~\cite{Erdi2016,Thattai2002}, concluding that specific network topologies such as long cascades with weak interactions and particular types of feedback motifs may suppress fluctuations~\cite{Chepyala2016,2014-iyer-biswas-mixedP}. We briefly elaborate on this analysis below.

Following~\cite{Thattai2002}, consider a signaling cascade of $N$ species $s_i$ where the production of species $s_i$ depends only on $s_{i-1}$, each species undergoes self-degradation, and there is an additive source of independent noise. While the dynamics can depend non-linearly on the reactants, near steady state they can be linearized for fluctuations about steady state. Thus, near steady state, $s_{i}$'s production rate is linearly dependent on $s_{i-1}$'s fluctuation from steady state, with a differential amplification rate equal to $c_{i}$, while  $s_{i}$'s differential degradation rate is proportional to its own deviation from steady state, with a degradation rate of $\gamma_{i}$. A solution to this linearized analytically tractable problem (see~\cite{Thattai2002} for details) showed that as long as the differential amplification rates are less than the corresponding degradation rates (i.e., the timescales of propagation of signal along the cascade is slower than the timescales of achieving steady state), the noise in the output species, $s_N$, is linearly dependent on the noise in the input species, $s_{0}$, with a linear coefficient $\propto e^{-N/N_0}/\sqrt{N}$ that is exponentially suppressed by the length ($N$) of the signaling cascade. Here, $N_0$ sets the attenuation scale for the cascade length and is found to decrease (i.e., attenuation is faster) when differential amplification rates are lowered. Thus, reaction cascades with small amplification-to-degradation rate ratios in each step serve to suppress noise transmission in the cascade. This property is useful in designing good signaling networks as follows.

In general, the dependence of the $s_{i+1}$ production rate on $s_i$ can be non-linear, for all reactions in the cascade, leading to the existence of multiple stable fixed points of the system (i.e., steady states). The specific fixed point attained by the system in steady state, the ``response'' of the cascade, is determined by the value of $s_{0}$, i.e., the input ``signal''. An ideal threshold response is achieved by a noiseless system with two stable fixed points separated by a threshold, when the input signal crosses a threshold separating two basins of attraction for the two output steady states. It has been shown that when the dependence of $s_{i+1}$ on $s_i$ is ultrasensitive (i.e., more responsive than hyperbolic Michaelis-Menten kinetics), there is a robust condition that the differential amplification rates must be smaller than the degradation rates to produce a desirable sharp digital-like output response despite a noisy input, for appropriate cascade lengths~\cite{Thattai2002}.

\subsection{Benefits of noise}\label{sec:noiseBenefits}

There are notable benefits to the presence of noise in distinct trafficking processes. Inherent stochasticity in directed motor transport plays a crucial role in circumventing roadblocks due to microtubule-associated proteins~\cite{Mogre2020}. Some motors like kinesin-1 follow individual protofilaments; thus, the stochastic dissociation of single kinesin-1 motors upon encountering an obstacle enables a cargo bound to a team of such motors to effectively bypass the obstacle~\cite{Ferro2019}. Other motors like kinesin-2 and dynein frequently side-step to neighboring protofilaments due to stochasticity, allowing them to successfully bypass obstacles~\cite{Bertalan2015,Ferro2019}.

In search processes involving directed motion with intermittent search (Sec.~\ref{sec:intermittentSearch}), the presence of stochasticity plays a crucial role in effectively locating a target. When the target is not directly positioned along the motor pathway, the vesicle must stochastically detach from the pathway and rely on diffusion to find the target. At low noise levels, characterized by a low diffusivity, the vesicle takes significantly longer to reach the target (in fact, in the absence of any noise, the vesicle remains confined to the motor pathway and never successfully locates the target). At high noise levels, the vesicle detaches too frequently from the motor pathway, leading to prolonged search times. Thus, an intermediate noise level provides the optimum balance to efficiently find the target. Cells may have the ability to modulate the noise level by controlling the binding affinity of the vesicle to the motor pathway, in this way regulating the detachment rate of the vesicle to ensure an optimal level of stochasticity and thus facilitating timely and accurate target localization.

One of the two cisternal maturation pathways discussed in Sec.~\ref{sec:maturation}~\cite{Vagne2018a} requires stochasticity to drive the maturation process of the compartment, without which the compartment would perpetually be in steady state with mixed components due to balanced influx and outflux of the respective components.

Stochasticity also plays an important role in mixing. For a large number of particles undergoing diffusive motion and starting from a non-uniform density gradient, Eq.~\eqref{eq:diffusionFP} implies that the higher the coefficient of diffusion ($k_BT/\gamma$)---in other words, the noisier the individual particle's trajectory---the faster the density gradient evens out to the steady-state density. There is much incentive for cells to maintain a uniform homeostatic concentration of substances. Active diffusion plays a crucial role in facilitating faster mixing. In this process, experimentally observed diffusivities in intracellular transport processes are much greater than expected to arise from purely thermal fluctuations and sharply decrease in the absence of ATP~\cite{Guo2014,Gupta2017,Lin2016,Smelser2015}.

\section{CONCLUDING REMARKS}\label{sec:conclusions}

Scaling up analytically tractable stochastic models of intracellular transport to develop a general conceptual framework remains an ongoing quest~\cite{Bressloff2013}. However, the identification of hidden simplicities in emergent whole cell phenomenologies have proved a useful route to relate stochastic models to cell-level phenomena in other systems~\cite{joshi2023intergenerational,joshi2023cellular,joshi2023emergent,joshi2023nonmarkovian}, and may offer a useful framework to approach similar problems in the context of intracellular transport. The advent of a new suite of quantitative, dynamic live-cell imaging modalities, which includes the highly suitable light-sheet based techniques, in conjunction with novel data analysis techniques makes this an exciting line of inquiry to pursue at this time~\cite{2022-rajagopal}(Appendix B).

Finally, we note that the endosomal network is context-specific at various levels. For instance, for a given cell type, distinct cargoes show disparate intracellular itineraries such as EGF (degraded in lysosomes) and transferrin (recycled to the plasma membrane)~\cite{Bakker2017,Mayle2012}. This can be further tuned by features such as concentration; for instance, EGF at lower concentrations instead causes its cognate receptors to be recycled~\cite{Sigismund2008}. Trafficking of a given cargo may depend on the cellular state; for example, receptor-bound insulin has been shown to display an altered balance of recycling to degradation in insulin target tissues in obese mice~\cite{Ryu2014}. Furthermore, distinct cell fates in distinct tissues or in developmental contexts may result from~\cite{2022-iyer-biswas-waddington} either changes in the stoichiometry of key molecules and protein isoforms or simply the morphology of the cell~\cite{Blue2018,York2020}. In pathophysiological contexts the endosomal system has also been shown to adopt a new homeostatic state, as both hijacking viruses and bacterial toxins can alter endosomal trafficking and acidification to promote infection. In these contexts, the interplay with energetic costs of cellular dynamics and metabolic homeostasis is likely to prove crucial~\cite{2021-iyer-biswas-bioenergetics}.

In conclusion, while we have gained a wealth of insights into the molecular and structural mechanisms of protein machineries, and have catalogued various behaviors of prominent cargoes and their destinations, significant gaps remain in our knowledge of how interactions, transport, and membrane remodeling come together in space and time to result in the beautiful choreography of robust cargo detection, trafficking, and specific delivery to targets that cells routinely perform.

\clearpage

\begin{summary}[SUMMARY POINTS]
\begin{enumerate}
\item
Commensurate with its importance to cellular function, intracellular transport is a complex, multiscale process that utilizes several physical transport modalities, alongside fine control of vesicular identity, to carry a variety of cargo to their destinations.
\item 
Stochasticity at the microscopic level is an intrinsic feature of intracellular transport, with noise arising due to physical transport mechanisms as well as biomolecular interactions.
\item
Despite the noisiness of constituent dynamics, precise and robust outcomes arise at the whole-cell level.
\item
Describing cell-level outcomes entails integration across disparate time-, length-, and abundance-scales, which in turn requires data spanning the scales of the relevant phenomena to inform parameter choices and guide development of models with testable hypotheses.
\item
 Advances in imaging methodologies can now be used to observe stochastic, dynamic processes with sufficient spatiotemporal resolution and statistical precision to relate specific molecular processes to cell-level outcomes. 
 \item
Given time series of the behaviors of all endosomes carrying a specific set of markers, variance-based measures of noise and FPT distributions are straightforward to calculate. These can be related to global outcomes, such as endpoints of specific maturation steps or internalization of cargo to target compartments.
\item
Eukaryotic cells seem to both suppress noise and also strategically use it to achieve desired outcomes, but clear-cut examples of each approach still remain largely anecdotal.

\end{enumerate}
\end{summary}

\begin{issues}[FUTURE ISSUES]
\begin{enumerate}
\item
While the significant differences in the intracellular organizations of terminally differentiated cells have been identified, it remains an open question as to how changes arising in the transport system support signaling during tissue patterning and differentiation. 
\item
In pathophysiological contexts, extending understanding to the dysregulation of trafficking of key biomolecules may provide a mechanistic approach for therapeutic drug  and nanoparticle delivery strategies, in turn enabling efficient release at specific subcellular locations.
\item Experiments measuring precise timing of cargo delivery are still lacking. Single particle tracking of various cargoes through the endosomal network will be a powerful approach to extract stochastic features of the system. 
\item A new conceptual framework for spatiotemporal organization of the intracellular transport network that yields quantitative principles that transcend system-specific details remains to be articulated.
\end{enumerate}
\end{issues}

\section*{DISCLOSURE STATEMENT}

The authors are not aware of any affiliations, memberships, funding, or financial holdings that might be perceived as affecting the objectivity of this review. 

\section*{ACKNOWLEDGMENTS}

S.A. thanks the National Health and Medical Research Council of Australia (APP1182212) and Monash Data Futures Institute Seed Grant. H.M.Y. is supported by an Australian Government Research Training (RTP) Scholarship. The EMBL Australia Partnership Laboratory (EMBL Australia) is supported by the National Collaborative Research Infrastructure Strategy of the Australian Government. K.J., R.R.B. and S.I.-B.  thank the Purdue Research Foundation and Purdue University start-up funds for financial support. K.J. and S.I.-B acknowledge support from the College of Science Dean's Special Fund, the Ross-Lynn Fellowship award and the Bilsland Dissertation Award (to K.J.).

%
\section*{LITERATURE CITED}

\appendix

\subsection*{Appendix A: Theoretical preliminaries}

\subsubsection*{Characterizing and quantifying `noise'}\label{sec:noiseMeasures}

Many biologically relevant quantities in the cells are characterized by random fluctuations around expected values and are thus stochastic variables~\cite{allen2010,bressloff2014}. Examples include the numbers of molecules, their spatial coordinates viewed both in static and dynamic contexts, and various inter-event time periods~\cite{bressloff2014,eling2019,tsimring2014}. Measures of noise serve to quantify deviations from deterministic behavior~\cite{tsimring2014}. Below we highlight common measures of noise that characterize various aspects of the width of the probability distribution of the relevant random variable.

\paragraph*{Absolute measure: variance and standard deviation.}\label{sec:variance}

The variance of a random variable $x$, with mean value $\langle x\rangle$, is defined thus: $\text{Var}(x) = \langle(x-\langle x\rangle)^2\rangle$. The angular brackets represent averaging of the enclosed quantity. The variance is an absolute measure of the square of the width of the probability distribution. Its square root, the standard deviation, $\text{SD}(x)=\sqrt{\text{Var}(x)}$, is thus an absolute measure of the width of the probability distribution and an absolute measure of noise. Such absolute measures are physically relevant in biological contexts involving the same kinds of random variables, such as when comparing levels of fluctuations in the timings of different events. Moreover, they are convenient to use due to the variance of the sum of independent random variables being equal to the sum of the variances of the individual random variables. Thus, the noise in each time step of a sequence of random processes, measured using their variances, can simply be summed to yield the time uncertainty of the overall process~\cite{Sanders2023}.

\paragraph*{Relative measure: Coefficient of Variation.}\label{sec:varianceToMeanSquared}

The canonical dimensionless measure for noise is the ratio of variance to mean squared, defined as the square of the ``Coefficient of Variation'', which describes the relative magnitude of fluctuations~\cite{everitt1998,Bruggeman2018}. This scale-independent measure allows comparisons across variables with different dimensions, including variables characterized by different length- or timescales. The Coefficient of Variation is used to directly characterize the width of the probability distribution as a fraction of the mean value. This measure is physically most meaningful when the random variable is non-negative~\cite{Bruggeman2018}.

\paragraph*{Measure of Poissonian character: Fano factor.}\label{sec:fanoFactor}

The Fano factor, defined as the ratio of the variance to the mean ($F_{x} = \text{Var}(x)/\le\la x\ri\ra$), is a useful measure of noise in stochastic dynamics of biochemicals since it provides insight into the type of process involved. Biochemicals governed by elementary chemical reactions, such as the simple birth-death process (Eq.~\ref{rxn:uniporterProduction} with $v_A=1$), have the steady-state Fano factor value of $F = 1$ corresponding to a Poisson distribution. A succession of such reactions typically results in an increase of the Fano factor beyond $1$ in the final product~\cite{2009-iyer-biswas,2014-iyer-biswas-mixedP}. In contrast, $F<1$ typically indicates the presence of negative feedback~\cite{2014-iyer-biswas-mixedP}.

\subsubsection*{First passage processes.}\label{sec:firstPassageTime}

In many biological processes, the first passage time (FPT) provides a useful framework for modeling stochasticity, especially in the context of the timing of events. The FPT refers to the duration required for a specific event to occur for the first time, starting from a well-defined initial condition. Examples include a randomly diffusing particle reaching a target location after starting some distance away, or the concentration of a particular biochemical species surpassing a critical threshold value after starting from some lower value~\cite{Gardiner2009,2016-iyer-biswas-FPT} (Fig.~\ref{fig:models}b).

The FPT problem can often be cast in the following universal formulation for solution by either analytic or numerical means. Consider a stochastic variable $x(t)$ that evolves according to specified rules, such as diffusion (see Eq.~\eqref{eq:diffusion} below) or stochastic exponential growth in the case of individual bacterial cell sizes between divisions~\cite{2016-iyer-biswas-FPT}. We  keep these definitions general, making no restrictive assumptions about the nature (for instance, the dimensionality) of $x$. We then wish to find the FPT ($\tau$) distribution, $P(\tau)$, for $x$ to start from a certain starting point $x(0)$ and reach a region $E$. The starting point may even be distributed according to some initial probability distribution, with zero probability to lie inside $E$ for the FPT problem to be meaningful. We now consider the stochastic evolution of $x$ in the presence of absorbing boundary conditions (defined as where the probability distribution of $x$ vanishes) on the boundary of $E$, calculating the total rate, $J(t)$, of probability absorption by the region $E$. The FPT distribution is then simply equal to the calculated rate of absorption: $P(\tau) = J(\tau)$. A specific application of this formalism to 1D diffusion (Sec.~\ref{sec:diffusion}) is outlined below.

Consider a particle diffusing in one dimension with a background drift velocity $v_{d}$ starting at $x(0) = 0$. We wish to calculate the FPT distribution to arrive at $x=L$. This model is relevant for describing lattice diffusion of MCAK, a member of the kinesin-13 family, which performs a 1D diffusion to find the end of microtubules. Upon encountering the tip, MACK utilizes ATP to depolymerize the microtubules~\cite{2006-helenius}. Following the formalism described in Sec.~\ref{sec:firstPassageTime}, we first need to solve Eq.~\eqref{eq:diffusionFP} with $\vb{F}/\gamma \to v_{d}$ with the absorbing boundary condition $P(x=L, t)=0$ and initial condition $P(x,0) = \delta(x)$ (the Kronecker delta function at $x=0$). The solution to this problem is possible through the method of images, and the FPT distribution is just the probability current entering $x=L$~\cite{1915-schrodinger,2016-iyer-biswas-FPT},
 \begin{equation}\label{eq:FPT1Ddiffusion}
 P(\tau) = - D \partial_{x}P(x,\tau)|_{x=L} = \frac{L}{\sqrt{4\pi D \tau^{3}}} e^{-\frac{(L - v_{d}\tau)^{2}}{4D\tau}}.
\end{equation}

This solution, known as the inverse Gaussian distribution, displays interesting properties. In the absence of drift, the distribution behaves like a power law, $\sim \tau^{-3/2}$ at large times $\tau \gg L^{2}/D$. This distribution is so broad that even its mean diverges. There is a slowly decreasing probability that the particle has not reached $x=L$, since there is a large region for making excursions in the direction opposite to the target. However, $P(\tau)$ is still normalized so that the particle always eventually reaches the target.

The presence of drift dramatically changes this situation. If the particle drifts toward the target, the FPT is normalized to $1$ and at long times $\tau \gg L/v_{d}$, the FPT is exponentially suppressed. Thus, all moments of the distribution are well-defined and the particle reaches the target in finite time. If, on the other hand, the particle moves away from the target, the FPT is no longer normalized to $1$, which indicates that there is a finite probability that the particle never reaches the target. Both these long-time behaviors are consistent with common intuition of the FPT process.

A useful consequence of the above formulation of the FPT problem is that the FPT distribution of a particle to reach an exit, starting from some specified region, is the same as the distribution of transit times (where return through the exit is disallowed). Thus, the mean FPT is a measure of the inverse rate of transfer of biomolecules across the intervening space~\cite{2012-lagache,2017-lagache,2011-amoruso}.

Finally, a major simplification occurs in the formalism when the random variable evolves monotonically, such as via a stochastic growth process without death, e.g., bacterial cell growth between divisions under balanced growth conditions~\cite{2014-iyer-biswas-PNAS,2017-iyer-biswas-intthresh}. Then, if the time evolution of the probability distribution of the random variable is known, the FPT distribution for hitting a threshold value is simply the negative time derivative of the cumulative probability distribution evaluated at the threshold~\cite{2016-iyer-biswas-FPT,2014-iyer-biswas-PRL}.

\subsubsection*{Emergence of directed motor motion from energy-consuming stochastic dynamics.}

The Langevin equation corresponding to motion in a given conformal state is given by the 1D version of Eq.~\eqref{eq:diffusion} with $\vb{F}(\vb{x})$ replaced by $-V'(x)$~\cite{Bressloff2013,Reimann2002}. Denoting by $w_{i\to j}(x)$ the position-dependent $L$-periodic transition rate from $i\to j$, the Fokker-Planck equation for the probability density in $i^\text{th}$ state evolves as follows (we have used $D_{i}\gamma_{i} = k_{B}T$):

\begin{align}\label{eq:ratchet}
    \partial_t p_i(x,t)=&\frac{1}{\gamma_{i}}\partial_x\left[\left(V_i'(x)+k_BT\partial_x\right)p_i(x,t)\right]+\sum_{j\neq i}\left[w_{i\to j}(x)p_j(x,t)-w_{j\to i}(x)p_i(x,t)\right].
\end{align}

This equation has been solved analytically for simple systems, such as those with two states and with a not unreasonable assumption $\gamma_{1}=\gamma_{2} = \gamma$~\cite{Julicher1997,Parmeggiani1999,Prost1994}. Can this system yield a net nonzero motor velocity? The average motor velocity for a two state system in steady state has been calculated to be~\cite{Bressloff2013},
\begin{equation}\label{eq:v2stateRatchet}
    v=\frac{Lk_BT}{\gamma}\left[\int_0^L\frac{\int_y^{y+L}e^{V(z)/k_BT}dz}{e^{V(y)/k_BT}-e^{V(y+L)/k_BT}}dy\right]^{-1},
\end{equation}
where,
\begin{equation}\label{eq:V2stateRatchet}
    V(x)=\int_0^x\left[\lambda(y)V_1'(y)+(1-\lambda(y))V_2'(y)\right]dy,
\end{equation}
and
\begin{equation}
    \lambda(x)=\frac{\sum_{n=-\infty}^\infty p_1(x+nL,t)}{\sum_{m=-\infty}^\infty \left[p_1(x+mL,t)+p_2(x+mL,t)\right]},
\end{equation}
where the right-hand side can be shown to be time-independent for generic transition rates. 

If the system is in thermodynamic equilibrium (with no energy input) so that detailed balance holds,
\begin{equation}
    \frac{w_{1\to2}(x)}{w_{2\to1}(x)}=e^{[V_1(x)-V_2(x)]/k_BT}=\frac{p_2(x)}{p_1(x)},
\end{equation}
and the value of $\lambda(x)$ can be shown to be,
\begin{equation}
    \lambda(x)=\frac{1}{1+e^{-[V_1(x)-V_2(x)]/k_BT}}.
\end{equation}
Since $V_1$ and $V_2$ are both periodic with period $L$, it follows that $\lambda$ is now also periodic with the same period, and hence from Eq.~\eqref{eq:V2stateRatchet}, $V$ is also periodic with period $L$. Thus, in Eq.~\eqref{eq:v2stateRatchet}, $e^{V(y)/k_BT}-e^{V(y+L)/k_BT}$ goes to zero, and hence the motor velocity $v$ is 0. 

Thus, analysis of this simple problem confirms the second law of thermodynamics, which mandates that maintaining a non-zero motor velocity requires maintaining energy input (in the form of ATP), which keeps the system out of detailed balance and hence avoids true thermodynamic equilibrium. How the energy input acts in a realistic scenario depends on the specifics of the reaction scheme for the ATP-related processes. Obtaining steady-state solutions to models incorporating such schemes generally necessitates numerical calculations such as shown in~\cite{Parmeggiani1999}. From the results of these calculations biologically relevant measures such as noise and efficiency (the mechanical work done per unit of energy consumed) can be evaluated~\cite{Bressloff2013,Julicher1997}. More complex generalized ratchet models involving multiple states have also been explored~\cite{Liepelt2007,Lipowsky2005}.

\subsubsection*{The law of large numbers. } According to the law of large numbers,  the accumulated outcomes of many repeated independent trials of a random experiment are proportional to the probabilities of those outcomes. While this law does match the prevalent common sense understanding of probability, we now provide a simple quantitative argument. Using the additive properties of the mean and variance (Sec.~\ref{sec:variance}), we conclude that the sum of $N$ independent identical random variables is a random variable whose mean and variance both are respectively $N$ times the mean and variance of the a single variable. Thus, it follows that the standard deviation is only $\sqrt{N}$ times the standard deviation of a single variable. Thus, the fractional width of the probability distribution of the sum of these variables (or any other random variable that is proportional to the sum, such as the mean of the random variables) is suppressed by a factor of $\sqrt{N}$ compared to the corresponding measure for a single variable. In other words, as $N$ becomes large, the sum of the independent random variables becomes more and more deterministic, when viewed as a fraction of their expected magnitude. When this property is extended to random variables counting the success of a given outcome in repeated independent trials as either $1$ (success) or $0$ (failure), we find that the accumulated outcome is deterministic after a large number of trials and is proportional to the probability of that outcome, thus proving the law.

\subsection*{Appendix B: Quantitative imaging and analysis}\label{sec:imaging}

Fluorescence microscopy of live cells is essential to the study of intracellular transport. Fluorescence recovery after photobleaching (FRAP), in which fluorescent molecules in a small region of interest are photobleached prior to imaging of the movement of unbleached molecules into the same region, permits calculation of molecular diffusivity~\cite{LippincottSchwartz2018}. Similarly, fluorescence correlation spectroscopy (FCS) enables calculation of diffusivities from the autocorrelation functions of fluorescence intensities in an illumination region of interest~\cite{Yu2021}. However, these ensemble-averaged methods are limited in the study of stochasticity in intracellular transport dynamics, where observations of single-vesicle dynamical data are necessary. Such results provide spatiotemporal information about the movements and interactions of specific sets of biomolecules, achieved by labeling these features of interest with specific molecular probes ~\cite{Zhang2021}.

Traditional epifluorescence microscopy yield such results, but with poor axial resolution and low signal-to-noise ratio due to out-of-plane illumination, confounding attempts to localize in three dimensions and leading to untenable levels of photobleaching over time. For this reason, other imaging modalities have been developed and adopted. Total internal reflection fluorescence (TIRF) provides a significant improvement in signal-to-noise ratio by exciting fluorescent molecules within $\sim$200 nm of the surface~\cite{Axelrod1981}. For example, it has been successfully used to observe the diffusive motion of molecules on the cell membrane~\cite{Axelrod2001}. Unfortunately, TIRF is limited to studying phenomena near the cell membrane; for this reason, light sheet microscopy (LSM) approaches that use selective illumination of a thin optical section have gained in popularity. As an example, lattice light-sheet microscopy (LLSM) employs an ultrathin light sheet to acquire images plane-by-plane to generate a three-dimensional (3D) volume, thus permitting rapid volumetric measurements of whole cells~\cite{Chen2014}. Importantly for the study of intracellular trafficking, because LLSM enables whole-cell volumetric imaging, it enables measuring all events within a prolonged duration of the measurement while simultaneously allowing measurement of fast dynamics (on the order of seconds) from high-resolution data ($\sim$220 nm in lateral, $\sim$320 nm in axial dimensions, and $\sim$2 s in time). Additionally, as this technique produces negligible photobleaching, specimens may be observed for $>$30 min. In principle, this permits the complete set of labeled molecules of interest (often consisting of two to four separately labeled species) within a whole cell to be visualized for extended periods of time, providing the observations required to interrogate the stochastic processes highlighted above, at multiple spatiotemporal scales.




\end{document}